\documentclass[letterpaper,12pt]{article}
\usepackage{graphicx}

\begin{document}

\title{New generation of neutron decay experiments }

\author{Vladimir Gudkov \\Department of Physics and Astronomy
\\ University of South Carolina \\
Columbia, SC 29208 \\ gudkov@sc.edu}


\maketitle

\begin{abstract}
New generation of neutron decay experiments, which can be done with high accuracy at new Spallation Neutron Sources, are very important for solution of current problems of fundamental physics and for tests of the Standard Model. The open problems of theory related to these experiments and possible approaches to their solutions are discussed. 

\end{abstract}

{\bf Kewords:} neutron, beta-decay, radiative corrections, standard model

\newpage

In tree approximation, and neglecting recoil corrections and electron polarization, the neutron decay rate has a simple representation \cite{gen1} in terms of  coefficients of angular correlations $a$,  $A$,  $B$ and $D$:  
 \begin{eqnarray}
\frac{d^3\Gamma}{dE_ed\Omega_ed\Omega_{\nu}}= \Phi (E_e)G_F^2
|V_{ud}|^2 (1+3\lambda^2)
\hskip 2cm \nonumber \\
\times (1+b\frac{m_e}{E_e}+a\frac{\overrightarrow{p_e}\cdot
\overrightarrow{p_{\nu}}}{E_e
E_{\nu}}+\overrightarrow{\sigma}[A\frac{\overrightarrow{p_e}}{E_e}+B\frac{\overrightarrow{p_{\nu}}}{E_{\nu}}+
D \frac{\overrightarrow{p_e}\times \overrightarrow{p_{\nu}}}{E_e
E_{\nu}} ]), \label{cor}
\end{eqnarray}
These coefficients depend only on one parameter  $\lambda $ -- the ratio of weak  axial-vector and vector nucleon coupling constants.   Here, 
 $\overrightarrow{\sigma}$ is neutron spin, $m_e$ is electron mass, $E_e$,
$E_{\nu}$, $\overrightarrow{p_e}$, and $\overrightarrow{p_{\nu}}$
are energy and momentum of electron and neutrino,  $G_F$ is Fermi
constant of weak interaction (obtained from the $\mu$-decay
rate), and $V_{ud}$ is the Cabibbo-Kobayashi-Maskawa (CKM) matrix
element. The $\Phi (E_e)$ function includes normalization
constants, phase-space factors, and standard Coulomb corrections.  The parameter $b$ is equal to zero for the standard vector -- axial vector type of weak interactions, and the parameter
 $D$ is related to time-odd correlations of spin and
momenta, therefore in the first Born approximation, it is defined by a time reversal violating process.

The  improving accuracy of neutron $\beta$-decay experiments results in the most precise measurements of the relative axial-vector coupling constant $\lambda$,  which is very important for many applications of the theory of weak interactions, including astrophysics,  since the star's neutrino production is proportional to $\lambda^2$. Even more important is the fact that precise measurement of neutron decay is related to the possibility of obtaining the CKM matrix element $V_{ud}$ in nuclear model independent way because neutron decay rate is proportional to the  $|V_{ud}|^2$. Therefore, further increase in the accuracy of experimental data in neutron decay  changes the status of these experiments and ranks them closely to the most important experiments in fundamental physics.  Precise measurements both for  neutron life time  and for angular coefficients provide the opportunity to extract hadronic vector weak interactions constant with the accuracy comparable to the $0^+ \rightarrow 0^+$ nuclear $\beta$-decay experiments. Therefore, one can solve the unitarity problem of the CKM-matrix. Currently, the best value of the first (and largest) matrix element $V_{ud}$ (for  $u$ and $d$ quark masses mixing) can be obtained from the measurement of nuclear Fermi transitions in $0^+ \rightarrow 0^+$ nuclear $\beta$-decay. However, the procedure of the extraction of this matrix element involves calculations of radiative corrections for Fermi transition in nuclei. Despite the fact that these calculations have been done with high precision (see \cite{townerh} and references therein), it is impossible to obtain the values of these corrections from independent experiments. The unitarity condition for the CKM matrix
\begin{equation}\label{unit}
   | V_{ud} |^2    +    | V_{us} |^2    +    | V_{ub} |^2   =  1
\end{equation}
gives the constraint on the sum of three matrix elements. Two of them, $V_{us} =0.2196\pm 0.0023$ and $V_{ub}=0.0036 \pm 0.0007$  \cite{pdg}, have been measured in high energy physics experiments (see also \cite{ckm03,sir03}). The first element $V_{ud}$ gives the dominant contribution to the unitarity equation and, therefore, it is crucial for the test of the Standard Model. The current value \cite{townerh} of the matrix element obtained from nuclear $0^+ \rightarrow 0^+$ nuclear $\beta$-decay is $0.9740 \pm0.0005$, the value obtained from neutron  $\beta$-decay is $0.9713\pm 0.0014$ \cite{abele}. One can calculate this matrix element from the unitarity condition constraint assuming that other matrix elements are known. It gives us $0.9756\pm 0.0004$ \cite{abele}.  Comparing these results, one does not see a statistically significant discrepancy in the values of the matrix element $V_{ud}$. Nevertheless, due to differences in these values, there is certainly room for new physics which can contribute to the unitarity equation at the level of $10^{-2} - 10^{-3}$ (see for example \cite{abele,holsttr,deutsch,hercz,yeroz,marc02} and references therein). To resolve this problem, the new generation of precision neutron decay experiments \cite{sns} are required.

To obtain the  parameter $V_{ud}$ from precise neutron decay data, one has to calculate all corrections for neutron decay with the appropriate accuracy. It is well known that recoil effects \cite{rec1,rec2} (see, also \cite{wilk1,gardner}) and radiative corrections \cite{sir2,sir5,garcia} essentially modify Eq.(\ref{cor}) and coefficients $a$, $A$, and $B$. These corrections became important at the level of few percents and should be carefully examined to produce the relevant background for the data analysis and to be able to search for new physics, because these corrections have the same order of magnitude as the expected deviations from the Standard Model. The main concern here is the accuracy and reliability of calculations of radiative corrections. They have been carefully calculated  for Fermi transitions, however, neutron decay contains contributions from Gamow-Teller transitions as well. Another problem is related to the traditional procedure of separation of radiative corrections into ``outer''  and ``inner''  parts. The first part is a universal function of electron energy and is independent of the details of strong interactions. The second part, dominated by the large QCD short-distance term, contains nucleon structure-dependent contributions. Since ``outer'' corrections could be calculated precisely, they do not bring uncertainties into extraction of parameters from experimental data. The problems arise with estimations of the ``inner'' corrections: this is because a part of them is very dependent on hadron structure model, and another (main) part results from QCD loop calculations with very high momentum transfer based on a consideration of the  nucleon as a system of free quarks. Moreover, the renormalization procedure \cite{sir6} mixes the leading ``outer'' and small ``inner'' parts of corrections and, as a consequence, leads to even more uncertain results (in terms of reliability and control).  

Let's for a time being neglect the necessity of the renormalization, and answer the following question: could we obtain (or restrict) the model dependent parts of radiative corrections from the complete set of neutron decay experiments?  One can measure at least four parameters with  high precision: total decay rate, $a$, $A$ and $B$ coefficients. Therefore, one could expect that simultaneous analysis of these data may lead to the over-defined system of algebraic equations with the possibility of extracting the unknown parts of radiative corrections. Unfortunately, the answer is: it is impossible in the standard framework even if we are to neglect the renormalization procedure. 

To show this, we use results \cite{sir2} of calculations of radiative corrections  in the first order of approximation in electromagnetic coupling constant $\alpha$, neglecting terms of order $\alpha (E_e/M)\ln (M/E_e)$ and $\alpha (q/M)$, where $M$ is a nucleon mass and $q$ is a transferred momentum.

As a rule, radiative corrections  affect the value and the Lorentz structure of the zeroth-order matrix element of neutron $\beta$-decay 
\begin{equation}
M_0=\frac{G_F|V_{ud}|}{\sqrt{2}}[\overline{u}_e (\gamma^{\alpha}+\gamma^{\alpha}\gamma_{5})v_{\nu}][
\overline{u}_p (\gamma_{\alpha}+\lambda \gamma_{\alpha}\gamma_{5}) u_n]. \label{matr0}
\end{equation}
However, it was shown \cite{sir2} that the strong interaction contributions of the radiative corrections do not change the Lorentz structure of the initial matrix element $M_0$ but rather renormalize the vector and axial-vector coupling constants in hadronic current by energy independent parameters $a_V$ and $a_A$, correspondingly  
\begin{equation}
M_{str}=\frac{\alpha}{2 \pi}\frac{G_F|V_{ud}|}{\sqrt{2}}[\overline{u}_e
(\gamma^{\alpha}+\gamma^{\alpha}\gamma_{5})v_{\nu}][ \overline{u}_p
(a_V\gamma_{\alpha}+a_A\lambda\gamma_{\alpha}\gamma_{5}) u_n].
\label{str}
\end{equation}
It should be noted that parameters $a_V$ and $a_A$ are dependent of details of strong interactions and that both of them include contributions from vector and axial-vector currents. The $a_V$ leads to an additional renormalization of the product of Fermi coupling constant and the CKM matrix element, and was a subject of intensive study (see for example, \cite{townerh,sir5,sir6} and references therein) for obtaining the CKM matrix element from $0^+ \rightarrow 0^+$ nuclear $\beta$-decay data. The parameter $a_A$ has not been studied since it does not contribute to $0^+ \rightarrow 0^+$ nuclear $\beta$-decay. 

At the level of the first approximation considered in the paper \cite{sir2}, one can show that the renormalization of the parameter $\lambda$ as
\begin{equation}
\lambda \rightarrow \lambda \frac{1+\frac{\alpha}{2 \pi}a_A}{1+\frac{\alpha}{2 \pi}a_V}
\end{equation}
 does not change either the form of the Eq.(\ref{cor}) or all the correlation coefficients, and that the parameter $(1+\frac{\alpha}{ \pi}a_V)$ is a common factor for Eq.(\ref{cor}). This means that in the given  approximation of the standard approach for calculations of radiative corrections, first,  one cannot obtain experimental restriction on the strong interaction dependent parts of radiative corrections and, second, it is impossible to obtain the non-renormalized parameter $\lambda$ from experimental data. This is true for any kind of neutron decay experiment.  

This result leads to the necessity of careful calculations of the hadronic model dependent parts of radiative corrections both for vector and axial-vector currents. The corrections for vector coupling constant  are necessary  for obtaining the CKM matrix element. The axial-vector corrections became important for studying neutrino nuclear interactions.    The knowledge of the parameter $\lambda$ with very good accuracy is in high demand from neutrino astrophysics. For example, the analysis of recent results from SNO experiment\cite{SNO} needs precise calculations\cite{NETAL,BCK,AETAL} of neutrino deuteron cross sections which are dominated by Gamow-Teller transitions. 

It should be noted that the above ``no-go'' result is  obtained in the first order of the approximation. In the second order, one can see that parameters $a_V$ and $a_A$ became dependent on electron energy. Therefore, when measuring neutron decay process with higher accuracy, it is possible to separate contributions from vector and axial-vector corrections and even to obtain restrictions on their values. Unfortunately, this approach  does not look realistic because the energy dependance of hadronic structure corrections arising in the second order of approximation is extremely small -- about $10^{-5}-10^{-6}$. 

What options do we have for control of the reliability of calculations of radiative corrections if we can neither obtain them from any set of neutron decay experiments nor calculate them in a model independent way using standard approach?   

One can suggest both experimental and theoretical solutions. The experimental approach is the comparison of precise experimental data of neutron decay experiments with elementary processes which are dominated by Fermi transitions (for example, $\pi$-meson decay) and Gamow-Teller transitions (neutrino deuteron reactions). For both cases the accuracy must be better than the one currently available. The theoretical approach is related to the possibility of avoiding hadronic model dependent contributions which are present in the conventional QCD based approach. From this point of view, the  effective field theory (EFT),  where the unknown high energy behavior can be integrated out and replaced by the set of low energy constants in the effective Lagrangian, looks very promising as an approach for a solution of this problem.   The payment for this is a number of unknown parameters (counter terms) that can be extracted from a set of independent experiments. The first result of calculations of the radiative corrections for neutron decay using EFT is presented in the paper \cite{eftcor}.

One can conclude that new generation of neutron decay experiments, which are being considered at new Spallation Neutron Sources \cite{sns}, are extremely important for understanding of currently unresolved problems of fundamental physics and for test of the Standard Model. This is also a challenge for theory. There are many questions to be answered and many problems to be solved for optimization of the future experiments and for unambiguous interpretation of experimental results.

\vskip 0.5cm {\bf ACKNOWLEDGEMENTS}

This work is supported  by the US Department of Energy, Grant No. DE-FG02-03ER46043.


\newpage

\begin{thebibliography}{99}

\bibitem{gen1}J. D. Jackson, S. B. Treiman and H. W. Wyld, Jr.,
Phys. Rev. {\bf 106}, 517 (1957).

\bibitem{townerh} I. S. Towner and J. C. Hardy, Phys. Rev. C {\bf 66}, 035501 (2002).

\bibitem{pdg}  Particle Data Croup, Phys. Rev. D {\bf 66}, 010001-113 (2002).

\bibitem{ckm03} G. Isidori, in Proc. on Workshop on the CKM Unitarity Triangle, IPPP Durham April 2003; arXiv:hep-ph/0311044.

\bibitem{sir03} A. Sirlin,
arXiv:hep-ph/0309187 (2003).


\bibitem{abele} H. Abele, NIM  A{\bf 440}, 499 (2000).


\bibitem{holsttr}  B. R. Holstein and S. B. Treiman, Phys. Rev. D{\bf 16}, 2369 (1977). 
 
\bibitem{deutsch} J. Deutsch, in: Fundamental Symmetries and Nuclear Structure, eds. J. N. Ginocchio  and S. P. Rosen,  p.36,World Scientific, 1989.

\bibitem{hercz} P. Herczeg,  in: Fundamental Physics with  Pulsed Neutron Beams , eds. C. R. Gould, G. L. Greene, F. Plasil and W. M. Snow, p. 64, World Scientific, Singapore, New  
     Jersey, London, Hong Kong,  2001.
     
 \bibitem{yeroz}B. G. Yerozolimsky, NIM  A{\bf 440}, 491 (2000).

\bibitem{marc02} W. J. Marciano,
Nucl. Phys. B (Proc. suppl.) {\bf 116}, 437 (2003).

\bibitem{sns} Proceedings of the Workshop on Fundamental Neutron Physics at the Spallation Neutron Source (FNPSNS 2001), Oak Ridge National Laboratory, Oak Ridge, TN, USA, (September 20 - 21, 2001) http://www.phy.ornl.gov/nuclear/neutrons/program.shtml. 



\bibitem{rec1}S. M. Bilen'kii, R. M. Ryndin, Ya. Smorodinskii and Ho Tso-Hsiu,
Sov. Phys. JEPT {\bf 37}, 1241 (1960).

\bibitem{rec2}B. R. Holstein,
Rev. Mod. Phys. {\bf 46}, 789 (1974).

\bibitem{wilk1}D. H. Wilkinson,
Nucl. Phys. A{\bf 377}, 474 (1982).

\bibitem{gardner}S. Gardner and C. Zhang, 
Phys. Rev. Lett. 86 (2001) 5666.


\bibitem{sir2} A. Sirlin,
Phys. Rev. {\bf 164}, 1767 (1967).
 
\bibitem{sir5} A. Sirlin,
Rev. Mod. Phys. {\bf 50}, 573 (1978).

\bibitem{garcia}A. Garc\'{i}a and M. Maya,
Phys. Rev. D 17 (1978) 1376.

\bibitem{sir6} W. J. Marciano and A. Sirlin,
Phys. Rev. Lett. {\bf 56}, 22 (1986).

\bibitem{SNO}
SNO Collaboration, Q.~R. Ahmad et al.,
Phys. Rev. Lett. 87 (2001) 071301;
ibid 89 (2002) 011301;
ibid 89 (2002) 011302.

\bibitem{NETAL}
S. Nakamura, T. Sato, V. Gudkov, K. Kubodera,
Phys. Rev. C 63 (2001) 034617;
S. Nakamura, T. Sato, S. Ando, T.-S. Park,
F. Myhrer, V. Gudkov, K. Kubodera,
Nucl. Phys. A 707 (2002) 561. 

\bibitem{BCK}
M. Butler, J.-W. Chen, 
Nucl. Phys. A 675 (2000) 575;
M. Butler, J.-W. Chen, X. Kong,
Phys. Rev. C 63 (2001) 035501.

\bibitem{AETAL}
S. Ando, Y.~H. Song, T.-S. Park, 
H.~W. Fearing, K. Kubodera,
Phys. Lett. B 555 (2003) 49.


\bibitem {eftcor} S. Ando, H.~W. Fearing, V. Gudkov,  K. Kubodera, F. Myhrer, S. Nakamura and T. Sato,
 arXiv:nucl-th/0402100 (2004).



\end{thebibliography}
\end{document}